\def\be{\begin{equation}}
\def\ee{\end{equation}}
\def\ba{\begin{eqnarray}}
\def\ea{\end{eqnarray}}

\def\slash{\hspace{-5pt}/}

\documentstyle[12pt]{article}
\begin{document}

\title{{\bf $e^+ \ e^-$ into 4 fermions + $\gamma$ with ALPHA}}

\author{
{\bf Francesco Caravaglios}\thanks{INFN Fellow},  
\\ {\it Department of Physics,
Theoretical Physics,
University of Oxford,}\\
{ \it 1 Keble Road,
Oxford OX1 3NP} \\ \it (e-mail: caravagl@thphys.ox.ac.uk) \\ 
{\bf Mauro Moretti,}
\\ \it (e-mail: moretti@vaxfe.fe.infn.it)
}
\date{}
\maketitle

\begin{abstract}
We  apply
the  iterative numerical algorithm ALPHA, which  automatically
generates 
the N-point Green
functions (as recently suggested \cite{alpha}) to
the computation of the production rate of four fermions 
and four fermions plus  photon in electron-positron
annihilation. The discussion of the physical results 
is preceded by an introduction on the algorithm.  
\end{abstract}

\vspace{-12cm}\rightline{ OUTP 9613P} 
\vspace{13cm}
\section{Introduction}
The center of mass energy available at the  LEP I\negthinspace{I}
collider 
will be well  above the 
threshold energy for W pairs 
productions
and one of the main goals 
of the collider is to allow the study of the W physics (as well
as of any new physical effect which might manifest 
at  the 200 GeV scale).
The expected integrated luminosity is  about $500 ~pb^{-1}$ per 
experiment    and this will allow   a relatively good measurement
of the W pair cross section and therefore of the weak boson  mass.
The W decays into two fermions and  thus the signal for such a process 
is the presence of four fermions with the appropriate charge (and flavor).   
The narrow width approximation namely the assumption that the rate is
given by the rate for W pair production times the branching ratio of
the W decay  receives important corrections
already at the tree level for two main reasons:
$i)$ the production is close to the threshold and the presence of a
lineshape for the W propagator significantly affects the rate,
$ii)$ other
graphs contribute to the
same final state in addition to those accounting for virtual W exchange.
Both correction are large enough to be 'detectable' with the expected 
statistics and
therefore a full computation is required.

We also present the calculation  of the production of four fermion
plus a  detectable photon
arising from anyone of the charged virtual/real particle involved in 
the process. 

To perform the computation of the cross sections we have  used the
code ALPHA  \cite{alpha} for the scattering matrix elements
 and the package VEGAS \cite{vega}  for the integration over  the phase space.
\subsection{The Method for the  Computation of the N-points
 Green Functions (ALPHA)}

In a recent paper, \cite{alpha},
we have suggested an {\it iterative} algorithm  
 to compute automatically  the scattering matrix elements 
of  any given effective
lagrangian, $\Gamma$. The $S$ matrix is recovered  from 
the effective lagrangian after  a functional  legendre transform.
By exploiting the  relation between 
$\Gamma$ 
 and the connected Green function generator, $Z$, we have obtained an
iterative
formula 
which   does not require the explicit use of the
Feynman graphs and it is suitable to be implemented in  a numerical  routine.
The problem of computing a N-point  scattering matrix element can be 
recast as the problem of finding the minimum of a function $\tilde
Z$ with 
respect to a finite set of variables. Once the subsequent 
 stationary conditions
for $\tilde Z$ are written down they can be solved iteratively (with respect
to  an expansion parameter). After a proper number of iteration steps one
obtains the scattering amplitude. 

The objects of interest to us are the connected Green Functions
of the theory $Z^{(n)}_{p_1\cdots p_n}=<\phi(p_1)\cdots\phi(p_n)>$.
 To compute one of them,  it is convenient to define
a function  $\tilde Z$ from the functional $Z$.
By means of a particularly simple choice of the source term
(following the notations in \cite{alpha})
\be 
J(x)=a_i e^{ - i p_i x}
\ee
we define 
\ba
\tilde Z & = & - \int \ d x \Gamma (\phi(x)) + \sum_{j=1}^n a_j 
\phi(p_j) \nonumber \\
\frac {\delta \tilde Z} {\delta \phi(p) } & = & 0 \label {stat} \\
\ea
where the stationary condition (\ref{stat}) implies that $\phi$
has to be considered as a function of the $a_j$ such to satisfy (\ref{stat}):
in other words if the field $\phi(x)$  is  written in terms of its 
fourier components
\be
\phi(x)=b_i e^{- i P_i x}
\ee
then the  $b_i$ are function of the $a_i$     since they satisfy the
equations $\partial \tilde Z/\partial b_i=0$.
The index {\it i} runs over a discrete  set of values. In fact,
at the stationary point (\ref{stat}),  only  the momenta 
$P_i$ which are a linear combination of the $p_i$ 
 with integer  coefficients have a
non zero $b_i$ (see after). 
Moreover  the Green Function $Z^{(n)}_{p_1\cdots p_n}$ is 
immediately found as
\be
Z^{(n)}_{p_1\cdots p_n} = \lim_{a_j \rightarrow 0}
\frac { \partial \tilde Z} {\partial a_1 \cdots
\partial a_n} 
\label{amplitude}
\ee
and  only a finite number of $b_i$ is relevant in  this
limit.
Indeed  the $b_i$  can be expanded in a powers series in the
$a_i$,
\be
b_j=\sum_{l_1=0,\cdots,l_n=0}^\infty b^{l_1,\cdots,l_n}_j 
a_{j_1}^{l_1} \cdots a_{j_n}^{l_n}
\ee
and all the terms  containing $a_i^n$  ($n>1$) do not contribute
 to the
limit (\ref{amplitude}).
 In practice  for a lagrangian  of the type 
\ba
{\cal L} & = &{1 \over 2} \phi^\alpha \phi^\beta
\tilde\Pi^{\alpha \beta} 
+{1\over 6} \phi^\beta \phi^\gamma\phi^\delta 
 {\cal O}^{\beta\gamma\delta} 
\ea
the stationary condition (\ref{stat}) will take the form 
\be
 a_i^\alpha={\tilde \Pi}_{im}^{\alpha\beta} b_m^\beta
 +{1\over 2} {{\cal O}_{ijk}}^{\alpha\beta\gamma} b_j^\beta b_k^\gamma.
\ee
where we have denoted with greek index all possible particles
degrees of freedom:
lorentz, color, \dots
This equation can be solved  perturbatively with respect 
to the $a_i$ or, equivalently, to the
interaction  terms ${\cal O}$: if we call   $b_{j,m}^\alpha$  the solution 
up to  order $m$, the following recursive 
relation holds\footnote{In the first step we have simplified 
a factor ${\tilde \Pi}_{im}^{\alpha\beta}$ to obtain the truncated 
Green function.}
                 
\ba
b_{j,0}^\alpha&=&a_j^\alpha~~j=1,n \nonumber \\
 b_{j,0}^\alpha & = & 0~~j>n
~~~~~~~~n= number ~of~external~particle \nonumber \\
  ...& & ... \nonumber \\
b_{j,m}^\alpha&=&-{1\over 2}\Pi_{j,t}^{\alpha \beta}
 {\cal O}_{t,k,l}^{\beta\gamma\delta} \sum_{r+s=m-1}
b_{k,r}^\gamma b_{l,s}^\delta \label {itstep} \nonumber \\
\label{iter}
\ea
and the amplitude is given 
by\footnote{One actually needs to know
a number of $b^\alpha_j $ lower than
the one which can be inferred from eqs.~(\ref{iter}) and (\ref{algorithm}) 
since the stationary condition can be used to further 
reduce the number of relevant variables. We refer
the interested reader to \cite{alpha}.}
\ba
{\cal A}_{p_1,...,p_n}&=&-{1\over 2}\sum_{s+r=n-2}b_{j,r}^\alpha
\tilde\Pi^{\alpha\beta}_{j,l}b_{l,s}^\beta 
-{1\over 6}
\sum_{s+r+t=n-3}{\cal O}_{j,k,l}^{\beta\gamma\delta} b_{j,r}^\beta
b_{k,s}^\gamma b_{l,t}^\delta +
b_{j,n-2}^\alpha a_l^\beta \tilde\Pi^{\alpha\beta}_{j,l}
\nonumber\\
\label{algorithm}
\ea
Each $b_{j,m}^\alpha$ has a precise physical meaning:
it represents the sub-amplitude  with {\it m} vertex interactions and with the
following external legs: one
external  leg  with off-shell momentum $P_i$ plus  any subset  
of  external on-shell momenta  chosen among the external momenta 
$p_i$. The $b_{j,m}$ is zero if there is no subdiagram with {\it m} vertex 
and with the required external legs.

Thus the above algorithm is a recursive relation between the
subdiagrams of the  full process:
all the off-shell subamplitudes  with $m$ vertex are given in terms 
of all the sub-amplitude with $k<m$ vertices.

It is manifest that this recursive formula does not require any symbolic
manipulation and 
does not require the knowledge of the  Feynman diagrams.
Once the numerical values of the external momenta $p_j$
are specified the $b_{j,r}^\beta$ and
${\cal O}_{j,k,l}^{\beta\gamma\delta}$ in 
(\ref{iter}) and (\ref{algorithm})
are just arrays of complex numbers and therefore this formula can be used
to implement a
 numerical code.

\subsubsection {The Code ALPHA}
When the initial and final states of the process are specified 
(type,  momenta and spin of the external particles)  
the program  ALPHA prepares an array $b_j^\alpha$ for all the possible 
degrees of freedom.

The coefficients
${\cal O}_{ijk}^{\alpha\beta\gamma}$ 
and $\tilde \Pi_{lm}^{\alpha\beta}$ are returned by some 
subroutines
as  a function of  the finite set of possible momenta $P_m$. 
   
The ALPHA code includes all the electroweak interactions and all 
the flavor content of the Standard Model (SM) 
and it can perform the computation of 
all 
electroweak matrix elements in the SM regardless of the initial
or final state type.
In addition, due to its simple logic, it 
allows to modify  the lagrangian 
with no excessive effort (by adding the proper subroutines
to compute the new $O_{ijk}$
interactions  and/or adding the relevant variables
for the new particles).

 The algorithm  defined from the recursive  relations
(\ref{algorithm}) is  fast  and the computing time 
of the code seems to  increase
roughly as $a^n$ (with $a\simeq 4$), 
$n$ being  the number of external particles, while 
the number of feynman graphs (and therefore the time for the
evaluation of their sum)    usually grows faster than $n!$.

The origin of  this difference might be  understood by 
 comparing the number of operations required in the evaluation of
the left-hand side and right-hand side of the equation
$a(b_1+\cdots+b_n)=a b_1 + \cdots a b_n$;
to compute the sums before the multiplications is  more  economical
than the inverse.
On the contrary  one uses to  compute all the Feynman graphs and only
at the very end  sums all their contributions.
However, especially for complicated processes,
the same sub-amplitude often occurs in several distinct Feynman graphs.
In the above algorithm this factorization is explicitly exploited
since the $b_{j,m}$ 
do represent such sub-amplitudes.

Finally, since the code is entirely numerical,  the output can be
immediately used for the integration procedure. We will comment
in the following sections about the reliability of the code
and its performances.

\subsection{The integration procedure} 
In order to construct a kinematically allowed momenta configuration
the phase space is factorized as a multiple decay process using the
formula \cite{part}
\be
d^{3 n-4} \Phi(P;q_1,q_2,q_3,...,q_n)=d^2 \Phi(Q=q_1+q_2;q_1,q_2)
d^{3 n -7} \Phi(P;Q,q_3,...,q_n) (2\pi)^3 d Q^2
\label{decomp}
\ee
whose main advantage is that one just need
a single subroutine for a two body decay and then by  repeated calls
to this subroutine it is straightforward
 to obtain an $N$ body phase space.

To compute the cross sections of interest one must overcome
some  problems \cite{weight}. When the number of phase space variables
is large the only possible approach to the phase space
integration is via MONTECARLO integration. 
The scattering amplitudes can be singular, or more
generally strongly peaked, in particular regions of the phase space.
This can be due, for example, to the presence of infrared (quasi-)
 singularities, or to the propagation of heavy unstable particles.
When this is the case some of the internal propagators become huge
and, as a result, the bulk of the cross section comes from
a very thin region of the phase space.
To deal with this problem it is mandatory to have a large  sample
of points in  the singular regions.
To this purpose we have used the package VEGAS \cite{vega}
whose underlying idea is to redefine
 the integration variables in
order to minimize the variance.
So, for example, for  the process $e^+ e^- \rightarrow
 \mu^+ \nu_\mu \tau^- \bar
\nu_\tau$ we have chosen as phase space variables
the invariant masses $\mu_1$ and $\mu_2$
of the fermion pairs $\mu^+ \nu_\mu$ and $\tau^- \bar
\nu_\tau$, the angle $\theta$ between the beam and the combined
$\tau^- \bar \nu_\tau$ momentum plus other five variables,
along the lines of (\ref{decomp}), which don't exhibit
peaking behavior. VEGAS then rescales properly $\mu_1$, $\mu_2$
and $\theta$ to account for the peaking behavior of the cross section,
namely it greatly increases the sampling for $\mu_1$ and $\mu_2$
close to the mass of the W boson and it increases the sampling
close to $\theta=0$.
The situation is more delicate when the number of singularities is bigger,
mainly when there are many almost on-shell virtual particles
which can lead to the same final state. This is the case, for
example, for the process 
$e^+ e^- \rightarrow \mu^+ \nu_\mu \mu^- \bar \nu_\mu$ 
which can proceed both via $W^+ W^-$ 
 and via $Z \ Z$ or $Z \ \gamma^*$
virtual production and
decay. VEGAS is efficient 
when the `(quasi-)singular' piece of the amplitudes
can be written as the product of  functions $f_j(x_j)$
(where $\partial f_j / \partial x_k \sim \delta_{jk} $)
of the  integration variables $x_j$, thus 
a simple choice of phase space variables along the lines
of (\ref{decomp}) is not satisfactory. To circumvent
the problem we have split the integration domain into two regions
and we have performed the integral with two different choices of
phase space variables namely:

$a)$ in the region where virtual W exchange is dominant 
we have used:
the invariant masses $\mu_1$ and $\mu_2$
of the fermion pairs $\mu^+ \nu_\mu$ and $\mu^- \bar \nu_\mu$ 
 the angle $\theta$ between the beam and the combined
$\mu^- \bar \nu_\mu$ momentum plus other five variables,

$b)$ in the region where virtual $Z$ or $\gamma$ exchange is
significant we have used:
the invariant masses $\mu_1$ and $\mu_2$
of the fermion pairs $\mu^+ \mu^-$ and $\nu_\mu \bar \nu_\mu$ 
 the angle $\theta$ between the beam and the combined
$\mu^- \mu^+$ momentum plus other five variables.

To precisely define 
 the two distinct regions we evaluate the inverse propagators
of the virtual $W^+$, $W^-$, $Z$ and $\gamma$
and we use the variables in $a$ when the product of the W propagators
is smaller than the one of the Z or of $Z$ and $\gamma$, otherwise we use
the variables defined in $b$.

An analogous problem occurs for the process
$e^+ e^- \rightarrow e^+ \nu_e \mu^- \bar \nu_\mu$. In this case
in addition to the usual W pairs exchange
an additional peak in the cross section occurs when
the final $e^+$ is emitted in the forward direction with respect
to initial one, thus allowing for a soft virtual photon
to be emitted. Again we split the integration
domain into two regions defined by the relative size
of the virtual $W^+$ and of the virtual $\gamma$ (invariant $e^+$, $e^+$
four momentum)
 propagator and we perform the integral using two different
phase spaces.

It is straightforward (only a bit tedious) to extend
the procedure to final states containing $e^+ e^-$ pairs
with the obvious drawbacks that the integration domain
needs to be split into a larger number of regions.

The processes with the emission of an additional final photon
have additional singularities. It is well known that
they are dominated by the emission of soft collinear photon and
this remain true even with the relatively hard cuts we impose
on the photon four momenta. One needs therefore a phase
space which allows to single out the soft and collinear
behavior of the cross section.
Let us consider the process 
$e^+ e^- \rightarrow \tau^+ \nu_\tau \mu^- \bar \nu_\mu \gamma$. 
The phase space is divided into three regions
according to which among the invariant masses of each charged fermions
with the photon is the smaller. If it happens to be the $e^+ \gamma$
or $e^- \gamma$ invariant mass than we choose as integration
variables the momentum of the photon its angle with respect to the beam
and the invariant $W^+$, $W^-$ masses and the angle of the W with
the beam. If the photon is more collinear with the $\tau^+$
we choose as variables the invariant masses $\mu^-\nu_\mu$,
$\tau^+ \nu_\tau \gamma$ and $\tau^+ \gamma$, then
the momentum of the photon and the angle between the $W^-$ and the beam.
Analogously if the photon is emitted closer to the $\mu^-$ (these
two last region are absolutely symmetric and the same VEGAS' sampling
can be used).

The procedure in other cases is completely similar.

\section{$e^+e^-$ Into Four Fermions (plus $\gamma$) }

   Before  computing the four fermion processes (plus $\gamma$)
we have to fix the  interactions in the  lagrangian and their couplings.
We consider the Standard Model lagrangian; 
the interactions involved in our processes 
include all the gauge bosons three-vertices, 
some four-vertices (for $4f+\gamma$)   
 and all these interactions are unambiguously defined once we require them to 
satisfy the $SU(2)\times U(1)$ 
symmetry conditions and we specify the couplings $g_2=0.651698$,
$g_1=0.357213$
 and the mixing angle 
angle $\theta_W={\arctan}(g_1/g_2)$  between the 
two neutral gauge boson.
The couplings among the gauge bosons and the fermions are chosen as follows:
the photon fermion interaction is settled to 
$\alpha(2 M_W)=e^2/(4 \pi)=1/128.07$,
the $W$-fermion interaction is given by the  $SU(2)$ coupling $g_2$.
Finally the $Z$-fermion vector $g_V$ and axial-vector $g_A$ coupling are 
derived from the SM relation $g_V/g_A=1-4 |Q| sin^2\theta_W$ 
($Q$ is the charge of the fermion 
in unit of  electron charge) and $g_A= g_2/2$.
Gluons are not taken into account.
The gauge boson masses (in GeV) are $M_Z=91.1888$ and $M_W=80.23$;
we added an imaginary  part to the masses in the propagators 
in order to take into account of the constant 
(or running) \cite{gauginv} width of the physical weak boson propagator
($\Gamma_W=2.4974$, $\Gamma_Z=2.03367$). 
All fermion masses are taken from the Review of Particle Properties
\cite{part}.
\subsection{Results and Comments}
 Whenever  possible the four fermion  processes are dominated
by the production of a W pair with the subsequent decay.
The naive narrow width approximation (i.e. to replace
the cross section by the cross section of the
process $e^+ e^- \rightarrow W^+ W^-$ times the branching ratio
of the W into the given channel) receives important corrections
already at the tree level due to the finite W width and
to the presence of non resonant diagrams leading to the same
final state as those mediated by the W decay.
The effect of the finite width is typically of several per cent
 and the effect of non resonant diagrams
is particularly important for processes where the
number of non resonant diagrams is high and other 
channels contribute (Z and soft virtual photon)
like for example $e^- \bar \nu_e \mu^+ \nu_\mu$
or $\mu^- \bar \nu_\mu \mu^+ \nu_\mu$ (and similar
semileptonic or hadronic channels).
Moreover the non resonant diagrams become more important
close to the threshold for W pair production 
namely into one  of the energy regions which
will be used for an accurate determination
of the W boson mass. Because of the high accuracy
aimed to at the LEP II experiments all these
effects need to be accounted for.
In table \ref{tff} we list the tree level cross sections
for several four fermion final states at a
c.m. energy of 190 GeV.
Within the Montecarlo error the results are in agreement
with those obtained by means of other methods [6]
thus proving the reliability of our code and its usefulness
to build up Montecarlo Event Generators for complicated
processes. It is important to stress here that the whole
procedure for the computation of the matrix element
is completely automatic; once the algorithm is implemented
(like in ALPHA for example)
the user  has only  to  specify the number and type
of external particles in an input file. As output  the code will directly give
the scattering matrix elements for any given spin and momenta configuration.  
In table \ref {tcpu} we report the CPU times required
for the event generation for ALPHA to assess the performance of our code
(although one should keep in mind that it is just the first
implementation of the algorithm and large room for optimization is
left).

The addition of a photon in the final state  introduces a factor 
$\alpha_{QED}$ in the total cross section, but the existence  of
infrared and collinear singularities, allows  large logarithms to
appear and  to give a sizeable cross section.

The most important diagrams to contribute are those
accounting for the emission of the photon from one of the 
external  charged fermions. 
In such a case the amplitude
gets a  factor
\be 
{i\over p\slash+k\slash-m }\epsilon\slash u(p)
\ee
coming from the internal propagator contiguous to the external photon. 
When $(p+ k)^2-m^2$  goes to zero, the amplitude increases.
This introduces
into the cross section $\sigma$
  a peak behavior with respect the angle $\theta$ (between
the photon and the charged fermion) and the energy $E_\gamma$ 
of the outcoming photon (neglecting suppressed  terms of the type
 $m/p\cdot k$)
\be
\sigma \sim {d E_\gamma d \cos\theta_\gamma \over E(1-\cos\theta_\gamma)}.
\ee
We have introduced a cut-off  to keep under control these singularities 
and to  take into account the experimental detector capabilities.
If we call $\theta_i$ the angle between the photon and the external
charged $i$~-~th fermion,
 then for each final state configuration, 
we define $\theta_\gamma$ as the minimum angle among the $\theta_i$.

In figures  (1-3) we plot the differential cross
section  for  different processes as function of $E_\gamma$ and
$\theta_\gamma$ after having  integrated  over all the other phase
space variables.
The corresponding cross sections are reported in table  \ref{tffg}.
We do report only a subset of the cross-sections for $e^+e^-$
into four fermion plus $\gamma$. The given processes, however, exhibit
the most general peaking structure for this class of processes
(with the exception of those containing an $e^+e^-$ pair 
into the final state) and therefore the extension
of the calculation to any other process of this type 
is immediate.

During  the LEP 2 phase these  processes will be observable and the
the hard photon emission  could  be measured. 
We have  checked that the overwhelming  contribution to these cross
sections  
is due 
to the emission of soft and/or collinear photons
 and that 
the four point gauge boson vertices are  negligible.
If the born cross section is multiplied by the eikonal factor
  (and the proper rescaling of couplings)
 one recovers with a relatively good approximation (from few to ten
percent in the relevant regions) the shown plots. 

In table \ref{tcpu} we report the CPU performances of the code
for a few four fermion plus $\gamma$ final states.
They are typically a factor three larger than
the corresponding ones for the processes without 
photon emission. In view of the large increase
in the number and in the complexity of the
Feynman graphs we regard this as remarkable
property of our algorithm.
\\
\\
\\

{\large \bf   Acknowledgments}\\
MM thanks the {\it Associazione per lo Sviluppo
della Fisica Subatomica,} {\it Ferrara} for financial support
and {\it INFN,} {\it sezione di Ferrara} for making available
computing facilities.

\eject

\begin{table}
\begin{center}
\begin{tabular}{|c|c|c|c|}
\hline \hline
process & cross section (pb) & process & cross section (pb) \\ \hline
$e^+ e^- \rightarrow
 \mu^+\nu_\mu \tau^- \bar\nu_\tau$ & 0.2185(1) 
&
$e^+ e^- \rightarrow 
\mu^+\nu_\mu \bar u d $ & 0.6663(4) 
\\ \hline
$e^+ e^- \rightarrow 
\bar u d \bar s c $ & 2.015(1) 
&
$e^+ e^- \rightarrow 
\bar e^- \nu_e \nu_\mu \mu^+ $ & 0.2271(1) 
\\ \hline
$e^+ e^- \rightarrow 
\bar e^- \nu_e u \bar d $ & 0.6929(5) 
& 
$e^+ e^- \rightarrow 
\mu^+ \mu^- \nu_\mu \bar \nu_\mu $ & 0.2286(2) 
\\ \hline
$e^+ e^- \rightarrow 
u \bar u d \bar d $ & 2.063(1) 
&
$e^+ e^- \rightarrow 
e^+ e^- \nu_e \bar \nu_e $ & 0.2573(2) 
\\ \hline
$e^+ e^- \rightarrow 
\mu^+ \mu^- \nu_\tau \bar \nu_\tau $ & 0.01010(1) 
&
$e^+ e^- \rightarrow 
\mu^+ \mu^- \tau^+ \tau^- $ & 0.00925(1)
\\ 
& &  & 0.01100(1) 
\\ \hline
$e^+ e^- \rightarrow 
\nu_\mu \bar \nu_\mu \nu_\tau \bar \nu_\tau $ & 0.008245(4) 
& 
$e^+ e^- \rightarrow 
\mu^+ \mu^- u \bar u $ & 0.02448(3) 
\\ \hline
$e^+ e^- \rightarrow 
\mu^+ \mu^- d \bar d $ & 0.02374(2) 
&
$e^+ e^- \rightarrow 
\nu_\mu \bar \nu_\mu u \bar u $ & 0.02104(2) 
\\ \hline 
$e^+ e^- \rightarrow 
\nu_\mu \bar \nu_\mu d \bar d $ & 0.01988(2) 
&
$e^+ e^- \rightarrow 
 u \bar u c \bar c$ & 0.05221(5) 
\\ \hline 
$e^+ e^- \rightarrow 
 u \bar u s \bar s$ & 0.04958(5) 
&
$e^+ e^- \rightarrow 
 d \bar d s \bar s$ & 0.04704(5) 
\\ \hline 
$e^+ e^- \rightarrow 
\nu_e \bar \nu_e \mu^+ \mu^- $ & 0.01778(2) 
&
$e^+ e^- \rightarrow 
\nu_e \bar \nu_e \nu_\mu \bar\nu_\mu $ & 0.008335(4)
\\ \hline
$e^+ e^- \rightarrow 
\nu_e \bar \nu_e u \bar u $ & 0.02389(2)
&
$e^+ e^- \rightarrow 
\nu_e \bar \nu_e d \bar d $ & 0.02066(2)
\\ \hline
$e^+ e^- \rightarrow 
\mu^+ \mu^- \mu^+ \mu^- $ & 0.005456(5)
&
$e^+ e^- \rightarrow 
\nu_\mu \bar \nu_\mu
\nu_\mu \bar \nu_\mu $
 & 0.004065(5)
\\ \hline
$e^+ e^- \rightarrow 
u \bar u u \bar u $
 & 0.02565(3)
&
$e^+ e^- \rightarrow 
d \bar d d  \bar d $
 & 0.02349(2)
\\ \hline
$e^+ e^- \rightarrow 
\nu_e \bar \nu_e \nu_e \bar \nu_e $
 & 0.004091(2)
\\ \hline
\end{tabular}
\end{center}
\caption {Cross sections
 in $pb$ for various 4 fermions final states in
$e^+ e^-$ collisions at a
center of mass energy of 190 GeV. The following cuts are applied:
The invariant mass of each $qq$, $q\bar q$ and $\bar q \bar q$ pair
is greater than 5 GeV, the energy of each final quark or antiquark
is greater than 3 GeV, the energy of each final charged lepton
is greater than 1 GeV, the angle of each charged leptons with
the beam direction is greater than $5^o$. 
The standard model couplings are defined in the test and
all fermion are massive with the masses as in [3].
For the final state $\mu^+\mu^-\tau^+\tau^-$
the cross section is given for both massive and massless fermion (lower row).
In all other cases the difference is small. Gluon exchange 
is not accounted for.}
\label {tff}
\end{table}

\begin{table}
\begin{center}
\begin{tabular}{|c|c|}
\hline\hline
 final state & $\sigma \;(fb)$  \\
\hline
 $\mu^+ \nu_\mu \tau^- \bar \nu_\tau \gamma $ & 22.652(46) \\
\hline
 $ e^+ \nu_e u^- d^- \gamma $ & 58.00(17) \\
\hline
  $ u^+ d^+ c^- s^- \gamma $ & 158.38(36) \\
\hline
  $ u^+ d^+ u^- d^- \gamma $ & 163.43(47) \\
\hline \hline
\end{tabular}
\end{center}
\caption { four-fermion + $\gamma$
final states in $e^+e^-$ collisions.
Inputs are the same as in table (1).
The cuts are the same as in table (1),
the photon energy is greater than 1 GeV, and the 
angle of the photon with the beam is greater than $11.5^o$ 
Gluon exchange 
is not accounted for.}
\label{tffg}
\end{table}

\begin{table}
\begin{center}
\begin{tabular}{|c|c|}
 \hline\hline
final state & CPU, sec/100000 events\\
\hline
 $ \mu^+ \nu_\mu \tau^- \bar \nu_\tau $ & 270 \\
\hline
 $ \mu^+ \nu_\mu \tau^- \bar \nu_\tau \gamma $ & 760 \\
\hline
 $ e^+ \nu_e u^- d^- $ & 445 \\
\hline
 $ e^+ \nu_e u^- d^- \gamma $ & 1388\\
\hline
  $ u^+ d^+ c^- s^- $ & 327 \\
\hline
  $ u^+ d^+ c^- s^- \gamma $ & 968 \\
\hline
  $u^+ u^- d^- d^+  $ & 605 \\
\hline
  $u^+ u^- d^- d^+ \gamma  $ & 1700 \\
\hline
  $e^+e^-e^+e^-   $ & 1120 \\
\hline
  $e^+e^-e^+e^- \gamma  $ & 3399 \\
\hline
  $e^+e^-e^+e^-  \gamma \gamma  $ & 10729  \\
\hline \hline
\end{tabular}
\end{center}
\caption 
{
Required CPU time for the computation (single precision)
 of 100000 
matrix elements for  various processes. 
All the Higgs couplings are set to zero.
The times are given in seconds
and the calculations have been performed with a DIGITAL machine
ALPHA 3000/600 with 64M of memory.
}
\label{tcpu}
\end{table}
\begin{figure}
\vskip 20truecm
\label{fig1}
\caption{differential cross section (fb/BIN) of 
 $ e^+ \nu_e u^- d^- \gamma $ as a function of  
$0.42<cos(\theta_\gamma)<0.98$ and $1~GeV<E_\gamma<19.5~GeV$
( see the text). All inputs parameters as in Table 1}
\end{figure}
 
\begin{figure}
\vskip 20truecm
\label{fig2}
\caption{differential cross section (fb/BIN) of 
 $ u^+ d^+ c^- s^- \gamma $
 as a function of  
$0.42<cos(\theta_\gamma)<0.98$ and $1~GeV<E_\gamma<19.5~GeV$
( see the text). All inputs parameters as in Table 1}
\end{figure}
 
\begin{figure}
\vskip 20truecm
\label{fig3}
\caption{differential cross section (fb/BIN) of 
  $ u^+ d^+ u^- d^- \gamma $ 
as a function of  
$0.42<cos(\theta_\gamma)<0.98$ and $1~GeV<E_\gamma<19.5~GeV$
( see the text). All inputs parameters as in Table 1}
\end{figure}

\end{document}